\documentclass[preprint,12pt,authoryear]{elsarticle}

\usepackage[markup=default]{changes}
\usepackage{amssymb}
\usepackage{amsmath}
\usepackage{algorithm}%
\usepackage{algorithmicx}%
\usepackage{algpseudocode}%

\usepackage{booktabs}%
\usepackage{multirow}%
\usepackage{graphicx}
\usepackage{svg}
\newtheorem{definition}{Definition}

\journal{Information Processing and Management}

\begin{document}

\begin{frontmatter}



\title{A Mechanistic Study on the Impact of Entity Degree Distribution in Open-World Link Prediction} %% Article title
%Relational Information-Driven Noise Robustness in Open-World Link Prediction

\author[organization]{Jiang Xiaobo} %% Author name
\ead{jiangxb@scut.edu.cn}
\author[organization]{Yongru Chen \corref{cor1}}
\ead{202220111978@mail.scut.edu.cn}
\cortext[cor1]{Corresponding author}
\affiliation[organization]{organization={School of Electronic and Information Engineering, South China University of Technology},%Department and Organization
            city={Guangzhou},
            postcode={510641}, 
            country={China}}

%% Abstract
\begin{abstract}
Open-world link prediction supports the knowledge representation and link prediction of new entities, enhancing the practical value of knowledge graphs in real-world applications. However, as research deepens, the performance improvements in open-world link prediction have gradually reached a bottleneck. Understanding its intrinsic impact mechanisms is crucial for identifying the key factors that limit performance, offering new theoretical insights and optimization strategies to overcome these bottlenecks. This study focuses on entity degree distribution, a core structural feature of knowledge graphs, and investigates its impact on the performance of open-world link prediction tasks. First, through experimental analysis, we confirm that entity degree distribution significantly affects link prediction model performance. Second, we reveal a strong positive correlation between entity degree and link prediction accuracy. Moreover, this study explores how entity degree influences embedding space distribution and weight updates during neural network training, uncovering the deeper mechanisms affecting open-world link prediction performance. The findings show that entity degree distribution has a significant impact on model training. By influencing the quality of the embedding space and weight updates, it indirectly affects the overall prediction performance of the model. In summary, this study not only highlights the critical role of entity degree distribution in open-world link prediction but also uncovers the intrinsic mechanisms through which it impacts model performance, providing valuable insights and directions for future research in this field.
\end{abstract}


%\item The impact mechanism of textual noise on open-world link prediction is revealed.
%\item A relation-enhanced method is proposed to improve noise resistance.
%\item A hierarchical gated fusion attention is designed to stabilize embedding distribution under noise.
%\item A relation clustering algorithm is proposed to construct relation-specific mapping function.

%% Keywords

\end{frontmatter}



\section{Introduction}
\label{sec1}
%% Labels are used to cross-reference an item using \ref command.

%%%%
Open-world link prediction methods expand the predictive boundaries of knowledge representation, enabling the knowledge representation and link prediction of new entities, thereby enhancing the practical utility of knowledge graphs in real-world applications. In recent years, many researchers have dedicated efforts to developing new models for open-world link prediction tasks, making notable progress in the field. These methods show immense potential in applications like question answering systems and recommendation systems, where they can improve the accuracy of inference \citep{ref_1,ref_2,ref_3}. Consequently, research in open-world link prediction holds significant academic and practical value.

At present, model optimization largely relies on trial-and-error approaches, where optimization methods are first proposed and then validated experimentally. While this approach has led to performance improvements, as research progresses, the gains in performance have gradually plateaued. These trial-based optimization strategies can no longer drive substantial improvements in open-world link prediction performance \citep{ref_4,ref_5}. Thus, in-depth analysis of the key factors influencing open-world link prediction performance, along with uncovering their intrinsic mechanisms, is crucial for providing new insights to overcome current performance bottlenecks. By exploring these intrinsic mechanisms, we can not only deepen our understanding of existing methods but also provide valuable insights for the development of more effective open-world link prediction models.

While studying these mechanisms is both necessary and significant, there is still insufficient research on how the characteristics of knowledge graphs influence link prediction performance at a deeper level. To date, the theoretical mechanisms behind these processes remain poorly understood and unexplored \citep{ref_6,ref_7}. Previous research has mainly focused on identifying factors affecting the quality of knowledge graph embeddings and link prediction performance, which can be categorized as significance analysis-based methods. For example, Lloyd et al. \citeyearpar{ref_8} examined the relationship between hyperparameters and the quality of knowledge graph embeddings, employing sensitivity analysis to evaluate the impact of adjusting hyperparameters and identifying the most significant ones. Yang et al. \citeyearpar{ref_7} performed a sensitivity analysis of factors affecting knowledge graph embedding learning, assessing how embedding algorithms, dataset structures, and training strategies impact embedding quality through link prediction outcomes. Although these studies offer valuable insights into the factors influencing knowledge graph embeddings and link prediction performance, they primarily focus on identifying these factors, and fail to explore the deeper mechanisms by which these factors impact model performance.

To address these limitations, this paper focuses on the key structural feature of knowledge graphs—entity degree distribution—and investigates its intrinsic impact mechanism on open-world link prediction models. First, we define the degree distribution index to quantify the imbalance in the degree distribution of knowledge graphs. Sobol sensitivity analyses and correlation analyses are then conducted to explore the relationship between degree distribution and model performance. Moreover, this paper delves into how entity degree affects embedding space distribution and weight updates during neural network training, revealing the deeper mechanisms by which it impacts open-world link prediction model performance. The experimental results confirm that entity degree distribution significantly influences link prediction model performance, with a strong positive correlation between entity degree and prediction accuracy. Mechanistic analysis shows that entity degree distribution impacts model performance during the training phase, affecting both the quality of the embedding space and weight updates, which in turn influences overall prediction performance. Finally, based on the findings from the mechanistic analysis, this paper discusses potential optimization methods, such as improved sampling strategies, data augmentation techniques, and adaptive loss functions, to enhance the model's link prediction performance. In conclusion, this study not only affirms the significant influence of degree distribution on open-world link prediction performance but also uncovers its intrinsic impact mechanisms. This research provides a mechanistic understanding for subsequent model improvements and offers new insights and directions for open-world link prediction.

%% Use \subsection commands to start a subsection.
\section{Related work}\label{sec2}
\subsection{Knowledge Graph embedding}\label{subsec2.1}
Knowledge Graph Embedding (KGE) aims to embed the entities and relations of a KG in a particular way into a low-dimensional continuous vector space \citep{ref_9,ref_10} . These low-dimensional vectors retain the inherent structural information in the KG, and the potential associations between entities and relations can be easily captured by measuring the low-dimensional vectors. KGE models can be classified into three categories based on the characteristics of the scoring function, namely distance-based models, tensor decomposition-based and neural network-based models. Representative distance-based models include TransE \citep{ref_11} , TransH model \citep{ref_12} , and various variants of them \citep{ref_13,ref_14,ref_15}  . Point-wise Euclidean space is the most commonly used representation space for them, projecting entity and relation embeddings in vector or matrix space \citep{ref_16}  . In addition, RotatE models \citep{ref_17} are represented in complex vector spaces, which define relations as rotations from source to target entities. Some distance-based models are modelled in manifold space, such as the TorusE model \citep{ref_18} . Models based on tensor decomposition include DistMult \citep{ref_19}  , ComplEx \citep{ref_20}, and others. Some KGE models are encoded based on deep neural networks. For example, some models use convolutional neural networks to extract features of entities and relationships in the knowledge graph, such as ConvE model \citep{ref_21}, SelectE model \citep{ref_24}. Some models use graph neural networks as encoders for capturing graph structure information, e.g., R-GCN model \citep{ref_22}, GAT \citep{ref_23}. The common feature of these KGE models is that they follow the closed-world assumption and can only perform closed-world link prediction tasks.
\subsection{Open-world link prediction}\label{subsec2.2}
For the KGE model, once training is complete, the embedding parameters are not updated and the distribution of the vectors remains fixed. For the representation of a new entity that has not appeared in the original knowledge graph, it is necessary to join the new entity and retrain the entire knowledge graph. The Open World Link Prediction (OW-LP) model aims to overcome the limitations of the KGE model by utilising the textual information of new entities to predict links for zero-shot entities \citep{ref_25,ref_26}. The mapping-based model transforms the textual embeddings of new entities into the knowledge graph embedding space, where they are scored and predicted using the scoring function of KGE. This class of methods has the advantages of lightweight, fast inference, and plug-and-play, and thus has a large potential for application. Sanh et al. first proposed OWE \citep{ref_27}, a mapping-based approach for open-world knowledge expansion, to expand the closed-world knowledge representation learning model. In recent years, a series of improved mapping-based open-world link prediction have been proposed \citep{ref_28,ref_29,ref_30,ref_31,ref_32}. For example, the EmReCo model \citep{ref_3} employs relational attention aggregation to acquire text features.The KRL\_MLCCL model \citep{ref_4} introduces the concept of comparative learning to the mapping-based OW-LP model. The mechanism study in this paper focuses on mapping-based open-world link prediction.

\subsection{Study of factors affecting performance}\label{subsec2.3}

Previous research works have been devoted to exploring the factors that affect the quality of knowledge graph embedding and link prediction performance. These works can be summarised as based on significance analysis. Lloyd et al. \citeyearpar{ref_8} used sensitivity analysis to investigate the relationship between hyperparameters and knowledge graph embedding quality. Yang et al. \citeyearpar{ref_7} conducted a sensitivity analysis on the factors affecting knowledge graph embedding learning and evaluated the significance of the impact of embedding algorithms, dataset structure, and training strategies on link prediction performance. However, these studies only scratch the surface of the influencing factors and do not reveal the intrinsic mechanisms by which these factors affect model performance. Several research works have discussed the impact of degree bias of nodes on model performance in graph learning tasks. Liu et al. \citeyearpar{ref_33} provided an overview of unbalanced learning on graphs and mentioned the impact of long-tailed distribution of node degrees on prediction performance. Subramonian et al. \citeyearpar{ref_34} explored the impact of degree bias in a node classification task for graph neural networks, describing potential mitigation routes.The work of Shomer et al \citeyearpar{ref_35} focuses on the phenomenon of degree imbalance in the link prediction task and proposes data augmentation methods to mitigate degree bias in knowledge graphs. However, the existing methods do not provide a good explanation for the mechanisms affecting degree distribution in open-world link prediction.

%% Use \subsubsection, \paragraph, \subparagraph commands to 
%% start 3rd, 4th and 5th level sections.
%% Refer following link for more details.
%% https://en.wikibooks.org/wiki/LaTeX/Document_Structure#Sectioning_commands

\section{Preliminaries}\label{sec3}

\subsection{Problem definition}\label{subsec3.1}

\begin{definition}[Knowledge Graph]
KG can be denoted by $\mathbf{G} = \{\mathcal{E},\mathcal{R},\mathcal{F}\} $, where $\mathcal{E}$ is the set of entities, $\mathcal{R}$ is the set of relations, and $\mathcal{F}$ is the set of factual triples. Binary concepts in KG are denoted by factual triple $(h,r,t) \in \mathcal{F}$, where $h$ is the head entity, $t$ is the tail entity, and $r$ is the relation between two entities, $h,t \in \mathcal{E}$, $r \in \mathcal{R}$. 
\end{definition}

\begin{definition}[Open-world Link Prediction]
Given a knowledge graph $\mathbf{G} = \{\mathcal{E},\mathcal{R},\mathcal{F}\} $, and an open-world entity set $\mathcal{E}_o$. Where the intersection of the open entity set $\mathcal{E}_o$ with the knowledge graph entity set $\mathcal{E}$ is empty, i.e., $\mathcal{E}_o \cap \mathcal{E} = \emptyset $. For the tail entity prediction task, given an incomplete triple $({e_o},r,?) \notin \mathcal{F}$, the goal is to accurately predict the missing tail entity $t$ in the triple, where $e_o \in \mathcal{E}_o$, $t \in \mathcal{E}$ and $r \in \mathcal{R}$. 
\end{definition}

\subsection{Methodology}\label{subsec3.2}
%% Inline mathematics is tagged between $ symbols.
This paper presents HGFA-OW, an open-world link prediction method we previously proposed, and details the experiments and data collection based on this method. Additionally, for comparison, the OWE model \citep{ref_27}, proposed by other researchers, is also included in the experiments.

Both models are grounded in the mapping paradigm, with the training process divided into two stages. The first stage involves knowledge graph embedding training, where translation-based and bilinear embedding models (such as TransE and Complex) are applied to train on closed-world knowledge graphs, thereby generating embedding vectors that represent the graph's structure. The second stage focuses on training the text-based embedding module and the mapping function. In open-world link prediction tasks, open-world entities (i.e., new entities) lack corresponding embedding vectors in the knowledge graph embedding space. Mapping-based models address this issue by expanding the embedding range of KGE models. Specifically, the model first generates text-based embeddings using the textual description of the new entity and then maps these embeddings into the KGE embedding space using the mapping function. In the second stage of training, the mapping function learns the alignment of entities between the text embedding space and the knowledge graph embedding space, completing the training process. Ultimately, open-world fact triples are scored in the KGE embedding space to predict links for unseen entities (open-world entities). 

\begin{figure}[htpb]
\centering
\includegraphics[width=\textwidth, trim=0.5cm 2cm 3cm 2cm,clip]{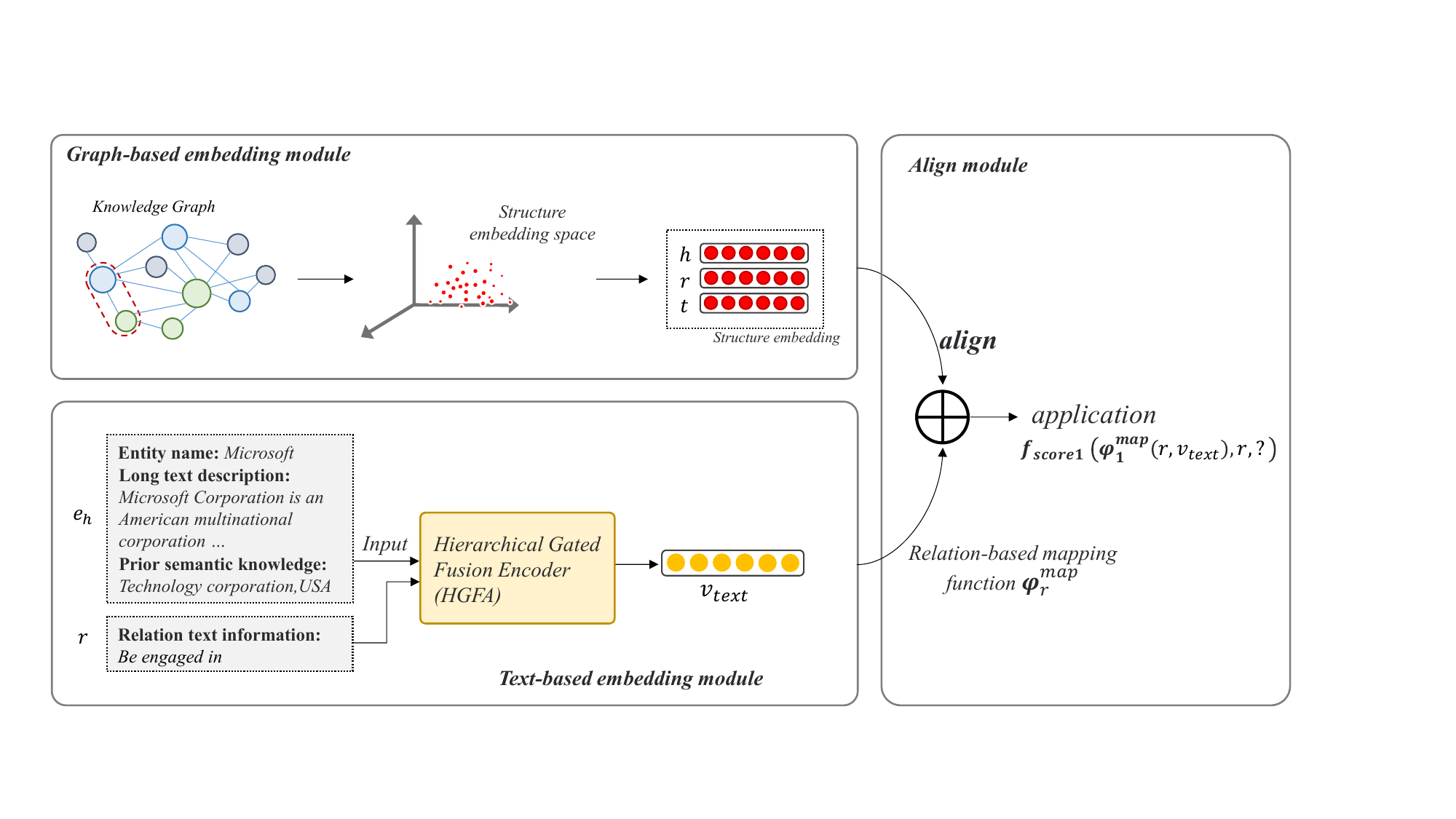}
\caption{Framework of the HGFA-OW model}\label{fig1}
\end{figure}

Detailed architecture of HGFA-OW model is shown in Fig.\ref{fig1}. The model can be divided into three parts, which are graph-based embedding module, text embedding module and alignment module. The text embedding module contains the hierarchical gated fusion attention encoder we designed. In the entity alignment module, we design a relationship clustering algorithm based on link similarity and semantic similarity to provide a basis for the classification of relationship categories. Independent neural networks are trained as mapping functions according to categories. In the inference phase, the textual embeddings of the open-world entities are mapped to the embedding space of the graph, and the candidate triples are scored and predicted in the KG embedding space.

In the first stage, for closed-world knowledge graph embedding training, we utilized the OpenKE training framework \citep{ref_36}. This framework integrates multiple mainstream knowledge graph embedding models and facilitates efficient model training and embedding vector generation. Using the OpenKE framework, we obtained graph-based embedding vectors, which were subsequently used as the foundation for training the second-stage open-world link prediction model.

\subsection{Dataset}\label{subsec3.3}
This paper conducts sampling, training, testing, and experimental analysis on the most commonly used open-world link prediction public dataset, FB15K-237-OWE. This dataset was constructed by Sanh based on the FB15K-237 dataset. Its test set contains 2081 open-world entities, which are used for the evaluation of the open-world link prediction task.
\subsection{Evaluation metrics}\label{subsec3.4}
The model's effectiveness was measured using two standard metrics: Mean Reciprocal Rank (MRR) and Hits@k (H@k). For each test sample, the correct tail entity is ranked according to the model's prediction results. The MRR is determined by averaging the reciprocal ranks across all test samples. Meanwhile, Hits@k quantifies the proportion of correct entities that fall within the top k predictions, with higher scores indicating improved predictive performance.

\begin{equation}
MRR = \frac{1}{{\left| {{D_{test}}} \right|}}\sum\limits_{{T_i} \in {D_{test}}} {\frac{1}{{ran{k_i}}}} \label{eq1}
\end{equation}
\begin{equation}
Hits@k = \frac{1}{{\left| {{D_{test}}} \right|}}\sum\limits_{{T_i} \in {D_{test}}} {[1 + I(ran{k_i} < k)]} \label{eq2}
\end{equation}

\section{Analysis of structural properties of KG}\label{sec4}
In this section, we define the structural characteristics of the Knowledge Graph that may be relevant to the performance of open-world link prediction and parameterise them for subsequent quantitative analysis. These relevant metrics are used to describe the structure of the knowledge graph dataset. In this paper, we propose a subgraph sampling approach to parameterise the structural features. Then the most critical global influences are identified through sobol sensitivity analysis. A correlation analysis is performed to assess potential linear associations between structural characteristics and model performance. In this paper, six metrics were finally selected as structural characteristics, and their definitions and descriptions are shown in Table 1.

\subsection{KG structural characteristics}\label{sec4.1}
Structural characteristics quantify the characteristics of the Open World Knowledge Graph dataset from a global perspective. These metrics are used to measure the overall linking of the entire graph, the distribution of nodes, and the complexity of the graph. In this section, we define the degree distribution index, which is used to measure the degree of imbalance in the distribution of node degrees. In addition, we define the relation type index, which is used to measure the degree of imbalance in the number of different relation types. The definition and computation of the indexes are described below.

\subsubsection*{1)	Degree distribution index}
We define the degree distribution index G to measure the degree of imbalance in the distribution of entity degrees in KG. The degree distribution index references the calculation of the Gini index \citep{ref_37}, where a higher index means a more pronounced entity degree imbalance. It is calculated as shown below:
\begin{equation}
    G = \frac{{\sum\nolimits_{i = 1}^n {\sum\nolimits_{j = 1}^n {\left| {{x_i} - {x_j}} \right|} } }}{{2{n^2}\bar x}}
\end{equation}
Where, ${x_i}$ is the number of linkage degrees of the \textit{i-}th entity, $\bar x$ is the average linkage degree, and $n$ is the number of nodes of the graph.

\subsubsection*{2) Relation type index}
Similar to the calculation of the degree distribution index, it is used to measure the degree of imbalance of relationship types in KG. A higher index means that the relationship types are more unbalanced, with certain types of relationships being in the majority. It is calculated as shown below:

\begin{equation}
    T = \frac{{\sum\nolimits_{i = 1}^n {\sum\nolimits_{j = 1}^n {\left| {{c_i} - {c_j}} \right|} } }}{{2{m^2}\bar c}}
\end{equation}
where $c_i$ is the number of samples that are in the $i$-th relationship category and $m$ is the number of relationship categories.

\begin{table}[htbp]
\centering
\caption{Definition of structural properties of KG}
\resizebox{\textwidth}{!}{  % 将表格缩放以适应页面宽度
\begin{tabular}{ll}  
\toprule  
Definition & Description \\  
\midrule  
Graph density & The ratio of the actual number of edges to the maximum possible number of edges \\
Global clustering coefficient & Reflect the degree to which the nodes in the graph form triangles \\
Number of relation types & Reflects the number of relation types in the sample graph \\
Degree distribution index & The degree of imbalance of entity linking degree in KG \\
Relation type index & The degree of imbalance in relation types in KG \\
Strongly connected components & The number of maximal subgraphs in which any two nodes $u$ and $v$ are path-connected in the graph \\
\bottomrule  
\end{tabular}
}
\end{table}

\subsection{Subgraph sampling method}\label{sec4.2}

To investigate the impact of structural characteristics on model performance, this study introduces a random subgraph sampling method. This approach leverages a Monte Carlo sampling-inspired strategy \citep{ref_38} to generate a set of subgraphs with varying structural characteristics, enabling the parameterization of structural properties in knowledge graphs. The sampled structural characteristics and their corresponding prediction performance metrics are subsequently utilized to analyze the sensitivity of prediction performance to structural characteristics. During the subgraph sampling process, initial nodes are selected based on a high-degree-first strategy, followed by breadth-first search (BFS) \citep{ref_39} to traverse nodes, ensuring the connectivity of the sampled subgraphs. Finally, the structural characteristics of the sampled subgraphs are computed and parameterized, with the detailed process outlined in Algorithm 1.

\begin{algorithm}[htpb]
\caption{Random subgraph sampling method}\label{algo1}
\begin{algorithmic}[1]
\Require Original knowledge graph dataset $G = (V,R,\xi )$, 
          Number of samples $n$
\Ensure The sampled connected subgraph ${G_{sub}}$, Structural characteristics of KG $\theta$
\State Load the original knowledge graph dataset $G$
\For{$i=1$ \textbf{to} $n$}
    \State $\triangleright$ The sampling ratio is chosen randomly
    \State $r \in random(0.5,1.0)$
    \State ${N_s} = round(\left| V \right| \times r)$
    \State $\triangleright$ Randomly select one of the k nodes with the highest link degree as the starting node
    \State ${S_k} = To{p_k}(),{v_0} \in {S_k}$
    \State $\triangleright$ BFS is used to sample the nodes to ensure the connectivity of the subgraph
    \State $S \leftarrow BFS({v_0},{N_s})$
    \State $\triangleright$ The set of sampled nodes and their edges form the sampled subgraph
    \State ${G_{sub}} \leftarrow (S,\{ (u,v)|u,v \in S{\text{ and }}(u,v) \in R\} )$
    \State $\triangleright$ The structural characteristics of the subgraph are counted
    \State $\theta  \leftarrow Statistic({G_{sub}}),\theta  \in {\mathbb{R}^{6 \times 1}}$
\EndFor
\end{algorithmic}
\end{algorithm}

\subsection{Sobol sensitivity analysis}\label{sec4.3}
To identify the structural characteristics that most significantly influence link prediction performance, we designed a Sobol sensitivity analysis experiment to evaluate the impact of these characteristics on prediction accuracy. The core concept of Sobol sensitivity analysis \citep{ref_40,ref_41} is to assess the contribution of each input variable (or its interactions with other variables) to the output variance by decomposing the variance into its constituent components. In this experiment, 50 connected subgraphs were randomly sampled, and their corresponding datasets were used for model training and testing. The Mean Reciprocal Rank (MRR) metric was obtained as the performance indicator for each sample. These labeled data points were then used to fit a regression model, which served as the basis for Sobol sensitivity analysis.

According to the variance decomposition framework, the variance of the model's output can be expressed as the sum of the variances contributed by individual input parameters and their interactions:
\begin{equation}
    \resizebox{\textwidth}{!}{
    $Var(Y) = \sum\limits_i^d {Var(f({X_i}))}  + \sum\limits_{1 \leqslant i < j \leqslant d} {Var(f({X_i},{X_j}))}  + ... + Var(f({X_i},...,{X_d}))$
    }
\end{equation}

The first-order Sobol index quantifies the contribution of a single input variable to the output variance and is computed as:

\begin{equation}
    {S_i} = \frac{{Var(f({X_i}))}}{{Var(Y)}}
\end{equation}

The second-order Sobol index captures the contribution of pairwise interactions between input variables to the output variance and is calculated as:

\begin{equation}
    {S_{ij}} = \frac{{Var(f({X_i},{X_j}))}}{{Var(Y)}}
\end{equation}

Higher-order Sobol indices are computed in a similar manner. Finally, the total Sobol index represents the overall contribution of an input variable, accounting for its individual effects as well as all higher-order interactions, to the output variance:

\begin{equation}
    {S_{Ti}} = {S_i} + \sum\limits_{j \ne i} {{S_{ij}} + } \sum\limits_{j \ne i \leqslant k} {{S_{ijk}} + } ...
\end{equation}

\begin{figure}[htpb]
\centering
\includegraphics[width=\textwidth, trim=0.5cm 0cm 0cm 0cm,clip]{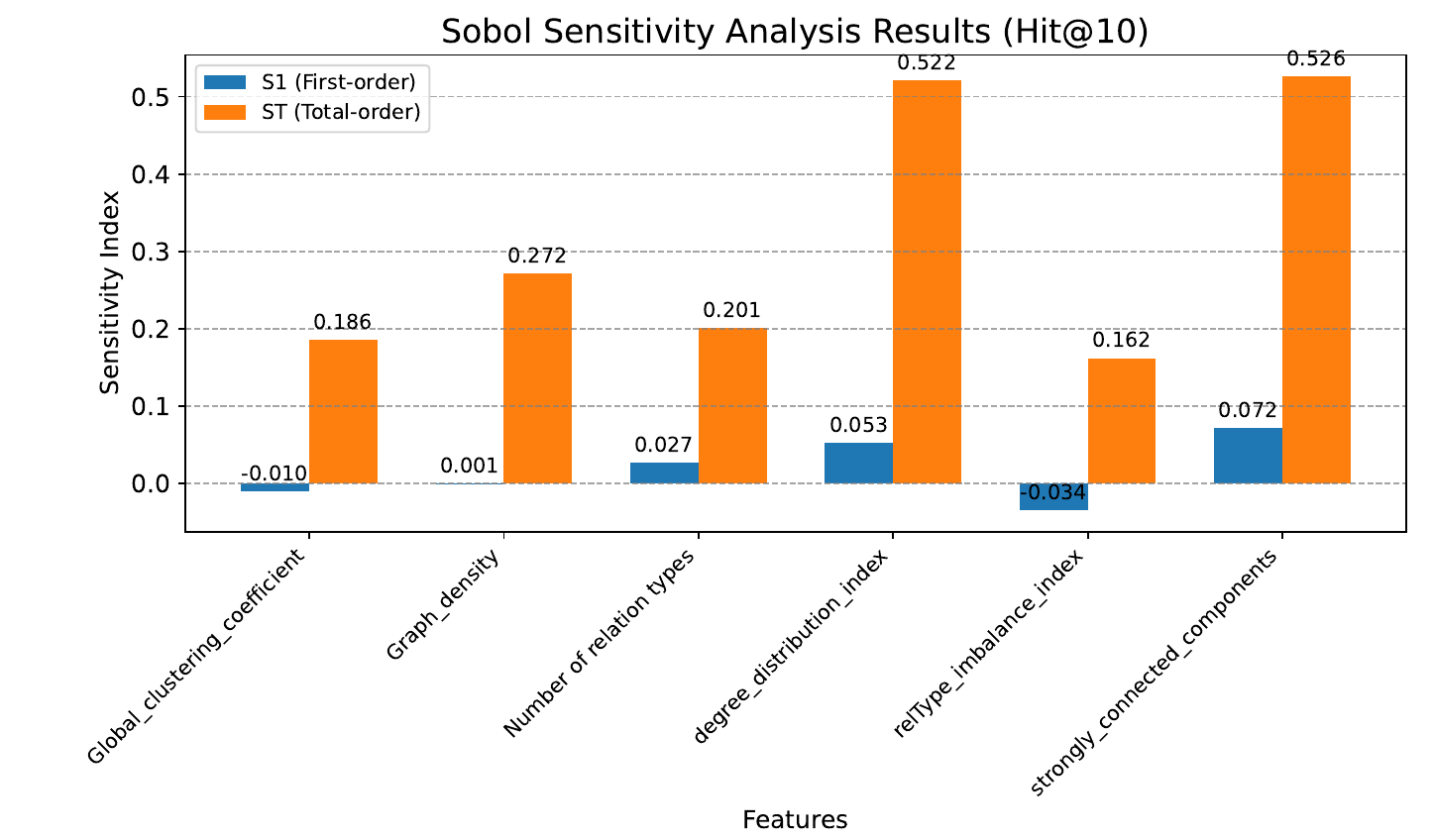}
\caption{Visualization of the sobol sensitivity index}\label{fig2}
\end{figure}

The results of the Sobol sensitivity analysis are illustrated in the figures. Figure \ref{fig2} shows the first-order and total-order Sobol indices for each structural characteristic, while Figure 3 presents the second-order Sobol index matrix as a heatmap. The experimental findings indicate that, among all the structural features, the degree distribution exponent and the strongly connected component have notably high first-order and total-order Sobol indices. This suggests that these two structural characteristics significantly influence link prediction performance. As observed in the second-order sensitivity index network in Figure 4, the interactions between the degree distribution index, strongly connected component, and graph density demonstrate high second-order Sobol sensitivity. This indicates that the degree distribution exponent not only has a direct effect on link prediction performance but also exerts an influence on the model's prediction results through its interactions with other structural features, such as the strongly connected component and graph density.

\begin{figure}[htpb]
\centering
\resizebox{0.8\textwidth}{!}{
\includegraphics[ width=\textwidth, trim=0cm 0cm 3cm 2cm,clip]{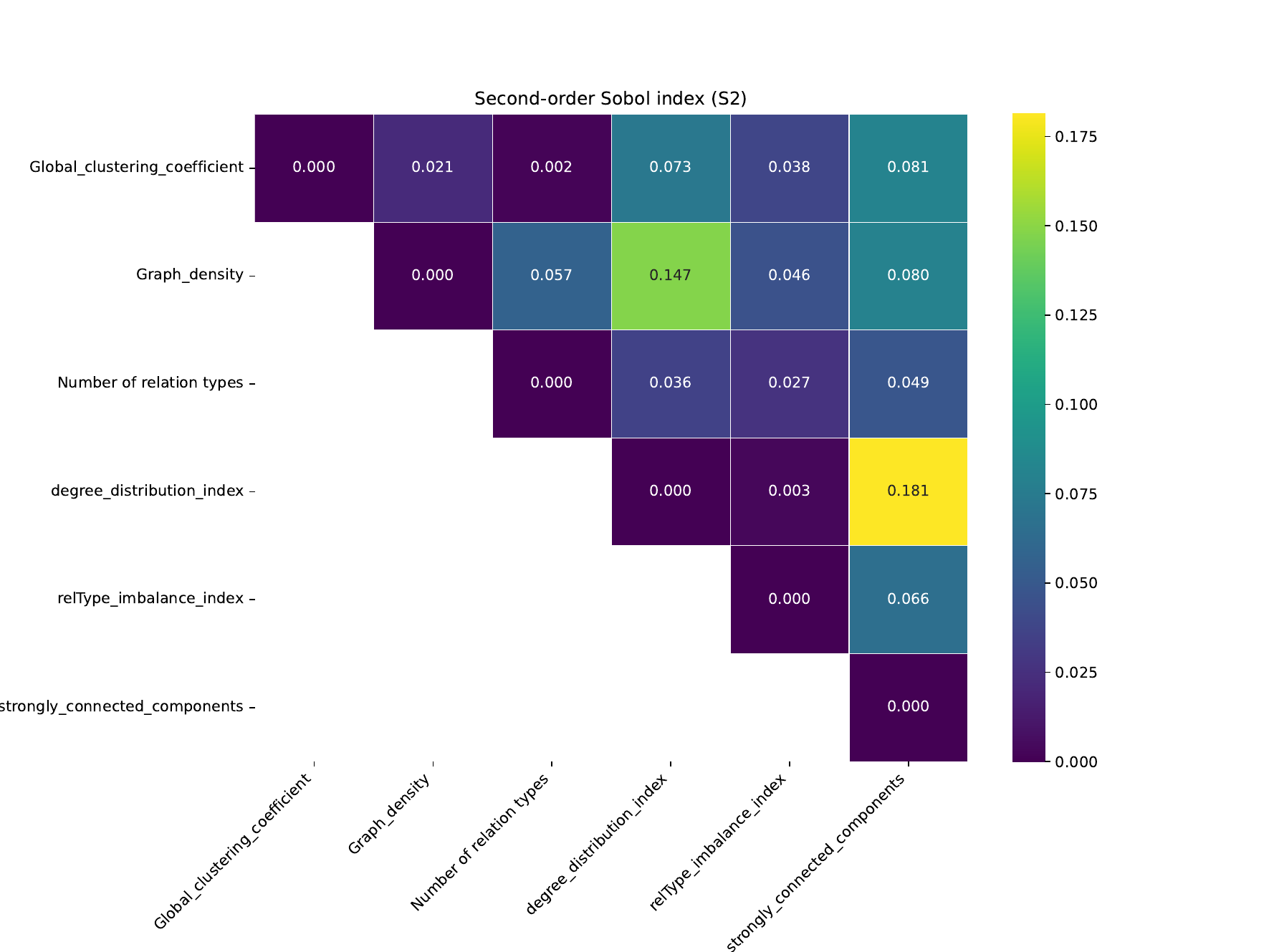}}
\caption{Heat map of the second-order sobol index}\label{fig3}
\end{figure}

\begin{figure}[htpb]
\centering
\resizebox{0.8\textwidth}{!}{
\includegraphics[width=\textwidth, trim=0cm 0cm 0cm 0cm,clip]{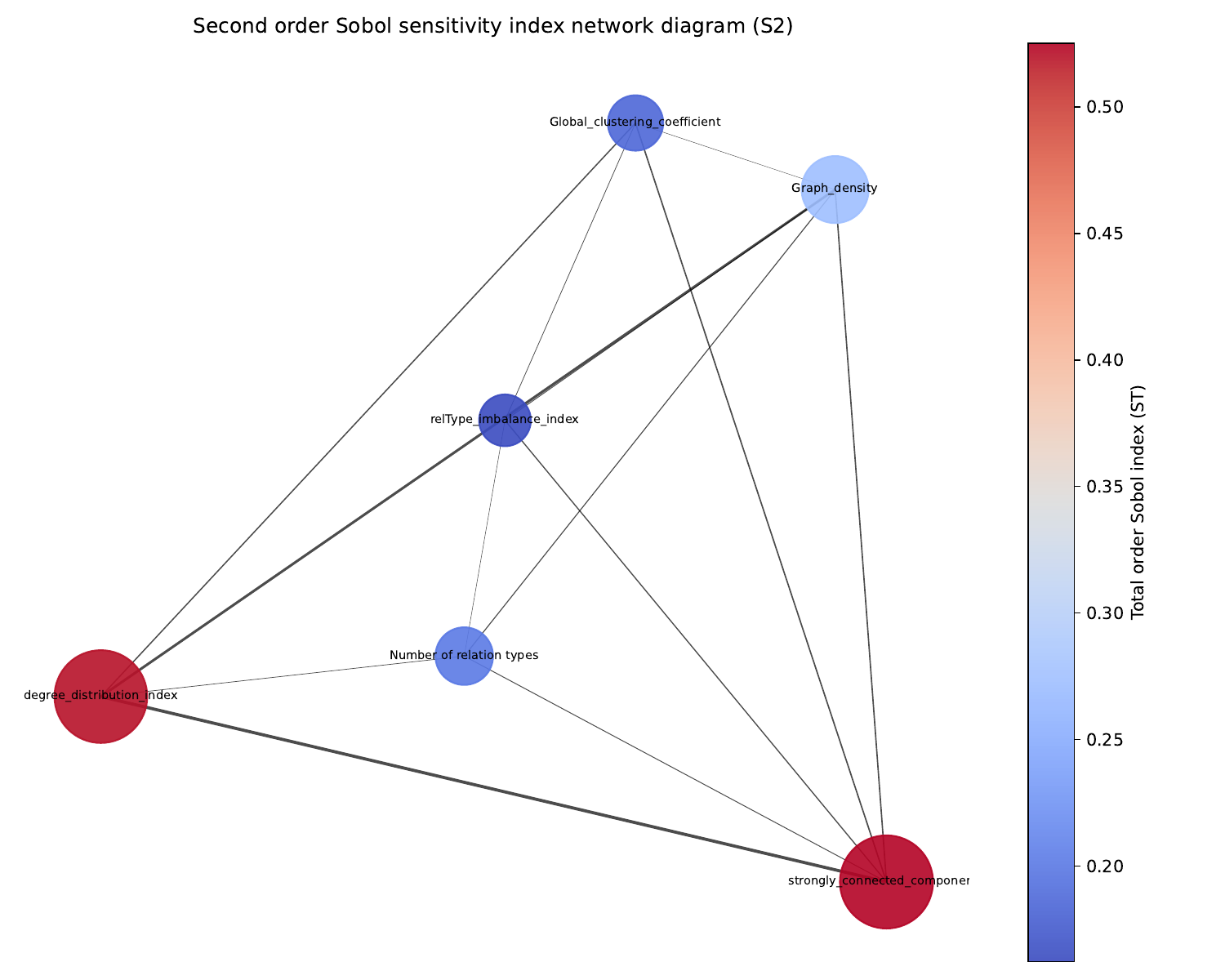}}
\caption{Network diagram of total and second-order sobol index}\label{fig4}
\end{figure}

\subsection{Correlation analysis}\label{sec4.4}
This study provides a systematic analysis of the correlation between structural characteristics and model performance. The distribution of sample points is visually displayed through scatter plots, and the relationships between structural characteristics and model performance are quantified using Spearman's rank correlation coefficient \citep{ref_42}  and Pearson's correlation coefficient \citep{ref_43}. This reveals the correlation between degree distribution and link prediction performance.

\begin{figure}[htpb]
\centering
\resizebox{\textwidth}{!}{
\includegraphics[width=\textwidth, trim=2cm 0cm 3cm 0cm,clip]{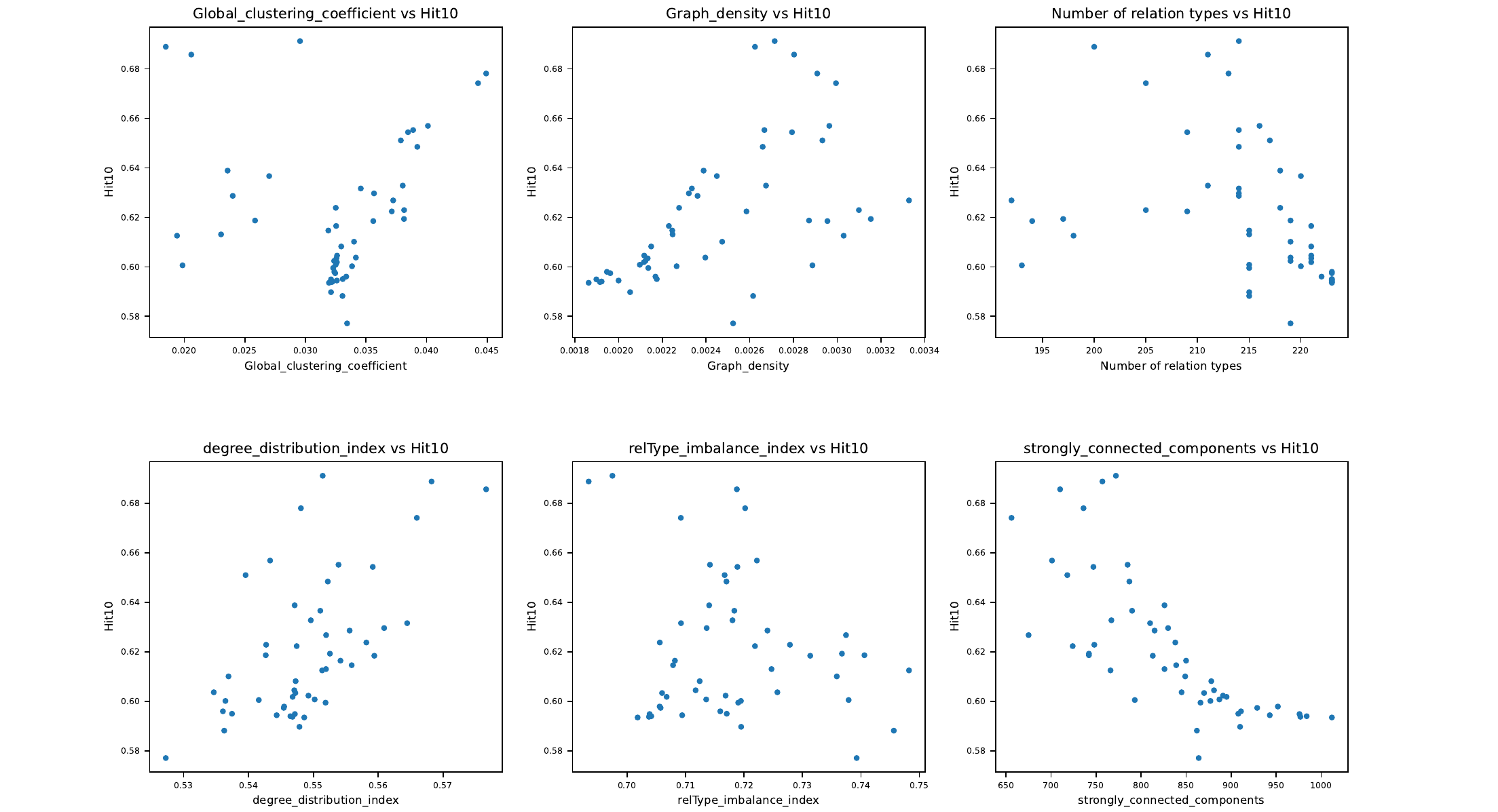}}
\caption{Scatter plot of the distribution of sample points}\label{fig5}
\end{figure}

\begin{table}[]
\centering
\caption{Correlation coefficient between degree distribution index and performance}
\resizebox{\textwidth}{!}{%
\begin{tabular}{llccc}
\hline
\multicolumn{2}{l}{\multirow{2}{*}{}}                                     & \multicolumn{2}{l}{Correlation coefficient} & \multirow{2}{*}{P\_value}    \\ \cline{3-4}
\multicolumn{2}{l}{}                                                      & MRR                 & Hit@10                &                              \\ \hline
\multicolumn{1}{c}{\multirow{2}{*}{Degree Distribution Index}} & Pearson  & 0.64                & 0.61                  & $5.8\times 10^{-7}$ \\ \cline{2-5} 
\multicolumn{1}{c}{}                                           & Spearman & 0.61                & 0.58                  & $2.4 \times 10^{-6}$ \\ \hline
\end{tabular}
}
\end{table}

The experimental results reveal a significant positive correlation between the degree distribution index and model performance, with Spearman's and Pearson's correlation coefficients of 0.61 and 0.64, respectively. As shown in the scatter plot, an increase in the degree distribution index corresponds to a noticeable improvement in the model's prediction performance. The degree distribution index is a metric used to measure the concentration of node connections within a knowledge graph. This finding highlights the significant impact of degree distribution on the performance of link prediction tasks.

\begin{figure}[htpb]
\centering
\resizebox{\textwidth}{!}{
\includegraphics[width=\textwidth, trim=3cm 2cm 3cm 2cm,clip]{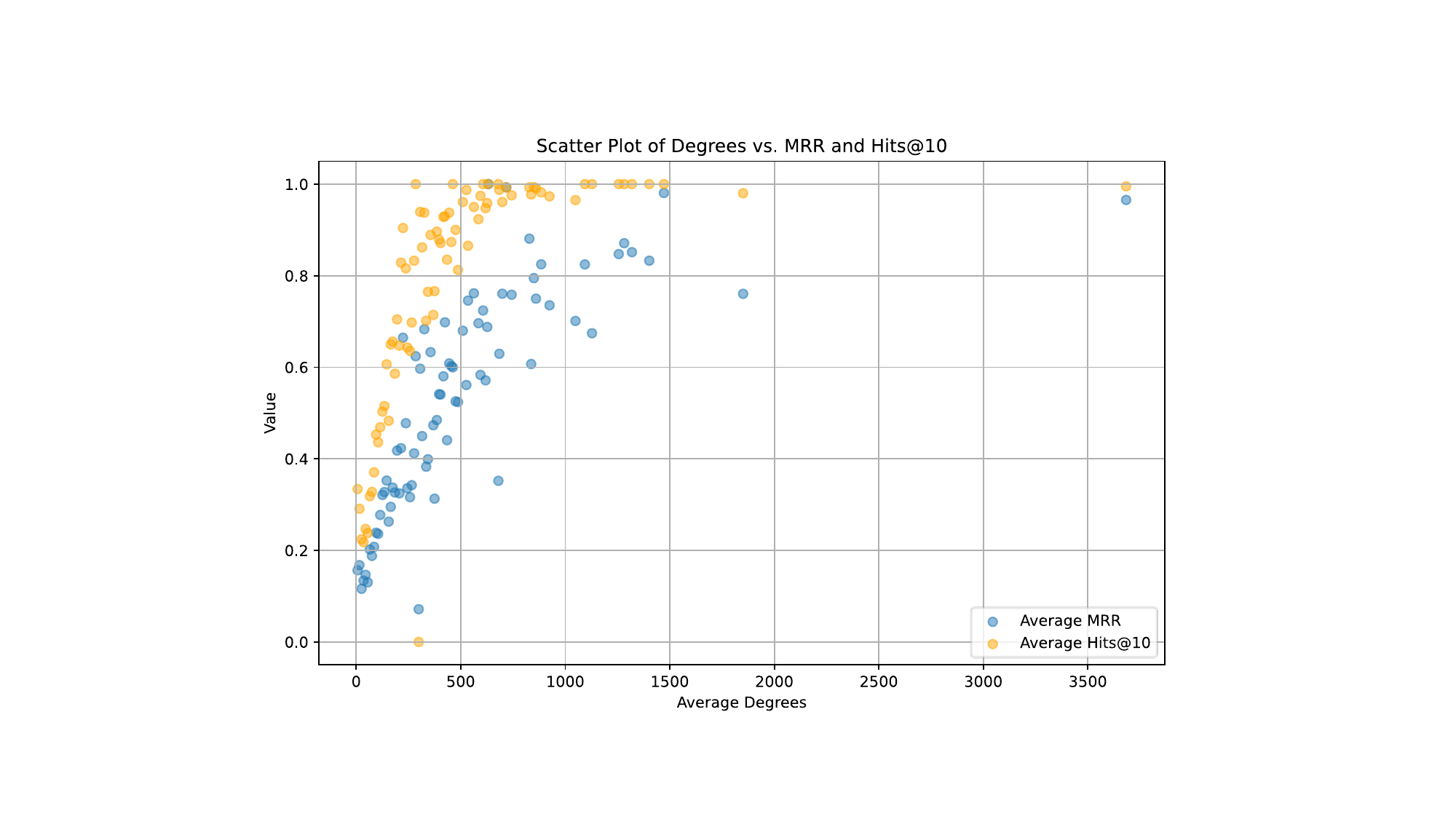}}
\caption{Scatter plot of the average entity degree versus the corresponding performance metrics}\label{fig6}
\end{figure}
Additionally, this study further investigates the relationship between entity degree and its prediction outcomes. The experimental results reveal a strong positive correlation between entity degree and the link prediction performance metric, MRR, with Spearman’s rank correlation coefficient and Pearson’s correlation coefficient reaching 0.89 and 0.71, respectively. This indicates that entities with higher degrees typically show superior link prediction capabilities. Specifically, by calculating the degree of each entity and comparing it to the prediction results associated with that entity, we found that high-degree entities generally exhibit more accurate prediction outcomes. This underscores the importance of entity degree not only within the graph structure but also in directly influencing the model's learning effectiveness for those entities.

\begin{table}[]
\centering
\caption{Correlation coefficient between entity degree and performance}
\label{tab:3}
\begin{tabular}{llccc}
\hline
\multicolumn{2}{l}{\multirow{2}{*}{}}                         & \multicolumn{2}{l}{Correlation coefficient} & \multirow{2}{*}{P\_value}    \\ \cline{3-4}
\multicolumn{2}{l}{}                                          & MRR                 & Hit@10                &                              \\ \hline
\multicolumn{1}{c}{\multirow{2}{*}{Entity degree}} & Pearson  & 0.71                & 0.57                  & $1.4\times10^{-13}$ \\ \cline{2-5} 
\multicolumn{1}{c}{}                               & Spearman & 0.89                & 0.90                  & $1.6\times10^{-28}$ \\ \hline
\end{tabular}
\end{table}

\section{Mechanistic analysis of the degree distribution effect on OW-LP performance}\label{sec5}
In Section 4, through quantitative analyses, we confirmed that the structural characteristic of entity degree distribution has a significant impact on the performance of link prediction models. In this section, we will investigate the intrinsic mechanism by which degree distribution affects link prediction performance in the context of open-world link prediction models. The influence of degree distribution on the internal training process of open-world link prediction models can be divided into two stages. The first stage is the effect of degree distribution on knowledge graph embedding training, including the effect on the distribution of embedding space and embedding quality. The second stage is the effect of entity degree distribution on the mapping function (neural network), including the alignment error of entities with different link degrees, and the degree of gradient contribution during the training process. In this paper, we design a corresponding analytical method to analyse the impact of degree distribution in the two phases of the model at the mechanism level.

\subsection{Effect of degree distribution on First-stage knowledge graph embedding training}
In the most commonly used mapping-based open-world link prediction models, the goal of first-stage knowledge graph embedding (KGE) training is to learn entity and relation representation vectors that encode the structural information of the graph. During this phase, the degree distribution of entities may influence the quality of the knowledge graph embedding vectors. To address this, we propose a knowledge graph embedding evaluation method and define relevant metrics to quantitatively assess embedding quality. This section outlines two analysis workflows: one for horizontally evaluating the impact of the degree distribution index on KGE pre-training quality, and another for vertically assessing how entity degree affects embedding quality.
Before the experiments, we define the criteria for evaluating knowledge graph embedding quality. KGE represents the knowledge graph in structural space, and entity embeddings contain the graph's link information. High-quality KGE should satisfy the following condition: entities with high link similarity should be positioned close to each other in the embedding space, whereas those with low link similarity should be farther apart. This evaluation criterion is analogous to the quality assessment in word embeddings, where semantically similar words are placed close together in the embedding space. Based on this standard, we define an embedding quality index Q, to quantitatively measure embedding quality. The specific definition process is detailed in Algorithm 2.

\begin{algorithm}[htpb]
\caption{Definition of the embedding quality index $Q$}\label{algo2}
\begin{algorithmic}[1]
\Require Knowledge graph dataset $G = (V,R,\xi )$, 
          Target entity ${e_0}$
\Ensure Embedding quality index $Q$
\State Load the KG dataset $G$, train the KGE model and obtain the embedding vectors $X$ 
\For{$e \in V$}
    \State $\triangleright$ Calculates link similarity with other entities based on
intersection and union ratio of neighbor entities and relations
    \State $Sim(e,{e_0}) = \alpha  \cdot Jaccard(N({e_0}),N(e)) + (1 - \alpha ) \cdot Jaccard(R({e_0}),R(e))$
\EndFor
\State $\triangleright$ Pick the $k$ most similar entity ${C_s}$ and the $k$ least similar entity ${C_d}$
\State ${C_s} = \arg \mathop {\max }\limits_{e \subseteq V,\left| C \right| = k} Sim(e,{e_0})$, 
\State ${C_d} = \arg \mathop {\min }\limits_{e \subseteq V,\left| C \right| = k} Sim(e,{e_0})$
\State $\triangleright$ Calculate the embedding quality index quantitatively
\State $Q({e_0},X) = \frac{{\sum\nolimits_{{e_j} \in {C_d}} {d(X[{e_0}],X[{e_j}]) - } \sum\nolimits_{{e_i} \in {C_s}} {d(X[{e_0}],X[{e_i}])} }}{{\sum\nolimits_{{e_j} \in {C_d}} {d(X[{e_0}],X[{e_j}])} }}$

\end{algorithmic}
\end{algorithm}

In this paper, we propose an analytical method for evaluating the effect of degree distribution indexes on KGE pre-training quality, as shown in Algorithm 3. The method reveals the influence of degree distribution on the embedding effect by quantitatively calculating the KGE training quality under different degree distribution indexes. Specifically, entities co-existing in the two subgraphs with high and low degree distribution indexes are extracted from the two subgraphs and traversed to compute their embedding quality indices in the embedding space of the two subgraphs, respectively. Ultimately, the method outputs the average embedding quality indices of the subgraphs with high and low degree distribution indexes as an assessment of the overall embedding space quality of the subgraphs.

\begin{algorithm}[htpb]
\caption{Analysis process of the impact of degree distribution index
on KGE pre-training}\label{algo3}
\begin{algorithmic}[1]
\Require High and low degree distribution index subgraphs ${G_1} = ({V_1},{R_1},{\xi _1}),{G_2} = ({V_2},{R_2},{\xi _2})$
\Ensure Average embedding quality index of two subgraphs ${\bar Q_1},{\bar Q_2}$
\State Load the subgraph datasets ${G_1},{G_2}$ , train the KGE model and
obtain the embedding vector ${X_1},{X_2}$ 

\State $\triangleright$ Extract the set of entities that co-exist in two subgraphs
${V_{common}} = {V_1} \cap {V_2}$

\For{${e_0} \in {V_{common}}$}
    \State $\triangleright$ The embedding quality index $q$ of the entity $e_o$ in the
two subgraphs is calculated by \textbf{Algorithm 1}, respectively
    \State ${q_{1i}} = Q({e_i},{X_1})$
    \State ${q_{2i}} = Q({e_i},{X_2})$

\EndFor
\State $\triangleright$ Calculate the average embedding quality index
\State ${\bar Q_1} = \frac{1}{{\left| {{V_{common}}} \right|}}\sum\limits_{i = 0}^n {{q_{1i}}}$, 
\State ${\bar Q_2} = \frac{1}{{\left| {{V_{common}}} \right|}}\sum\limits_{i = 0}^n {{q_{2i}}}$
\end{algorithmic}
\end{algorithm}

Furthermore, we examined the effect of entity linkage degree on the quality of KGE embeddings, as outlined in Algorithm 4. This is a longitudinal analysis intended to assess the differences in embedding quality between entities with different linkage degrees within the FB15K-237-OWE original knowledge graph dataset. First, based on the distribution of entity degrees in the dataset, we established thresholds to classify entities into high-degree and low-degree categories, thus creating sets of high-linkage and low-linkage entities. To ensure fairness in the experiment, we fixed the number of entity samples and then sampled an equal number of entities from each set. Next, we computed the embedding quality index for each sampled entity within the embedding space. Finally, we computed the average embedding quality for both high-linkage and low-linkage entities. This approach enables us to quantitatively evaluate the embedding quality differences between entities with varying linkage degrees.

\begin{figure}[htpb]
\centering
\resizebox{\textwidth}{!}{
\includegraphics[width=\textwidth, trim=6cm 5cm 7cm 3cm,clip]{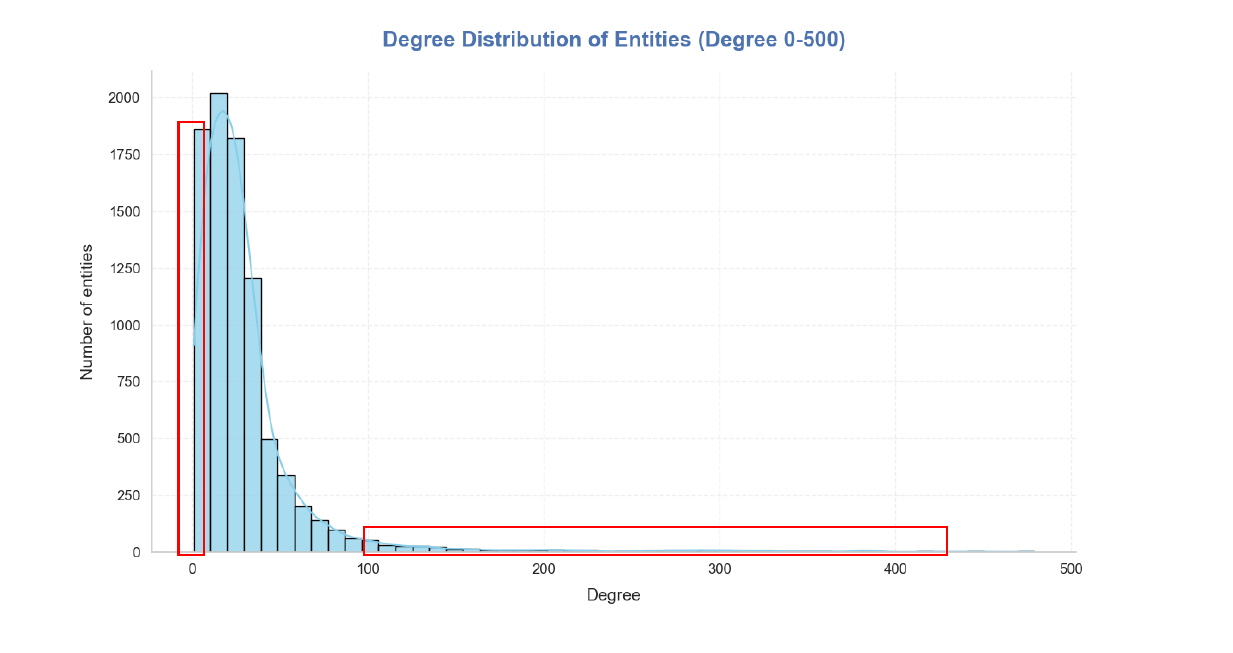}}
\caption{Entity degree distribution in FB15K237-OWE dataset}\label{fig7}
\end{figure}

\begin{algorithm}[htpb]
\caption{Analysis process of the impact of entity degree on KGE pre-training}\label{algo4}
\begin{algorithmic}[1]
\Require FB15K237-OW dataset $G = (V,R,\xi )$, Number of samples $n$, Partition threshold of high and low degree entities ${\varepsilon _H}$, ${\varepsilon _L}$
\Ensure Average embedding quality for entities with low and high linking degrees ${\bar Q_{low}},{\bar Q_{high}}$
\State Load the KG dataset $G$, train the KGE model and obtain the embedding vectors $X$ 

\State $\triangleright$ Calculate the degree of each entity in the graph $G$
\State $d(e) = \operatorname{degree} (e),\forall e \in V$
\State $\triangleright$ Divides the entity sets with high and low linking degree
\State $L = \{ e \in V|d(e) \leqslant {\varepsilon _L}\}$,$H = \{ e \in V|d(e) \leqslant {\varepsilon _H}\}$
\State $\triangleright$ Random sampling
\State ${H_s} = random\_sample(H,n)$,
\State ${L_s} = random\_sample(L,n)$
\For{${e_i} \in {L_s}$}
    \State $\triangleright$ The embedding quality index is calculated by \textbf{Algorithm 1}
    \State ${q_i} = Q({e_i},X)$
\EndFor
\For{${e_j} \in {H_s}$}
    \State ${q_j} = Q({e_j},X)$
\EndFor

\State $\triangleright$ Average embedding quality index for low and high degree entities
\State ${\bar Q_{low}} = \frac{1}{n}\sum\limits_{i = 0}^n {{q_i}}$, 
\State ${\bar Q_{high}} = \frac{1}{n}\sum\limits_{j = 0}^n {{q_j}}$
\end{algorithmic}
\end{algorithm}

\subsection{Effect of degree distribution on Two-stage mapping function training}

The training of the two-stage mapping function involves training a set of relationship-specific neural networks, with the main goal of aligning the text embedding space with the graph embedding space. This alignment enables the transformation of entities from the text embedding space to the KGE embedding space for OW-LP tasks. In knowledge graph datasets, the degree distribution of entities is directly related to the sampling of the training data. Consequently, entity degree distribution can affect model performance by influencing key intermediate variables during the training process of the mapping function (neural network), such as gradient norms, alignment errors, and loss values. To investigate the intrinsic role of entity degree distribution in this process, we designed corresponding experiments. Specifically, we categorized entities into high-degree and low-degree groups, recording the intermediate variables during training and monitoring the variations in gradient norms and alignment errors.

\begin{figure}[htpb]
\centering
\resizebox{\textwidth}{!}{
\includegraphics[width=\textwidth, trim=6cm 4cm 7cm 7cm,clip]{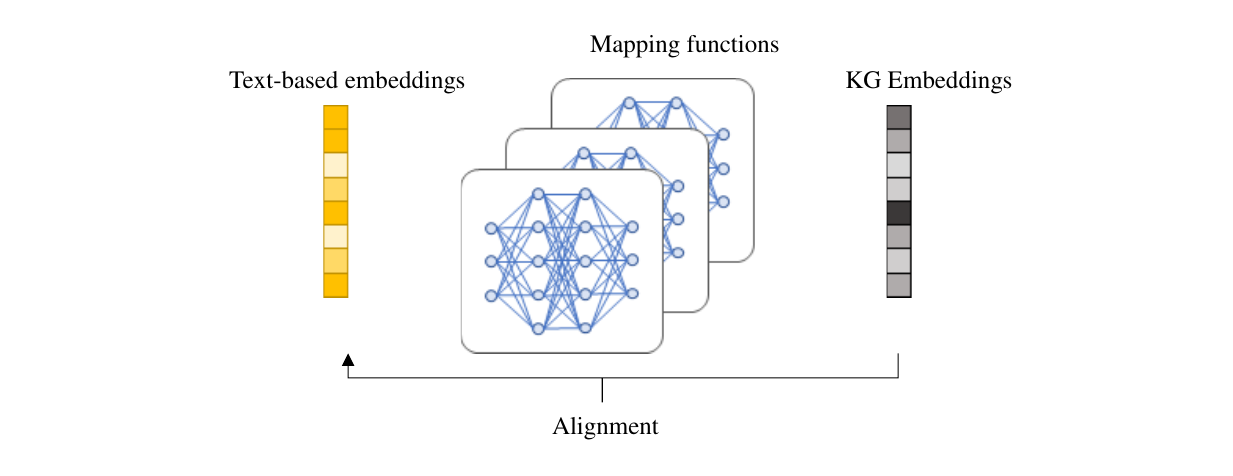}}
\caption{Diagram of the mapping alignment of the two stages}\label{fig8}
\end{figure}

\subsection{Results and Discussion}
Through the experiments and the analysis of the results in this section, we address the following questions:
\begin{itemize}
    \item \textbf{RQ1}: How does the degree distribution affect the training of open-world link prediction models in both stages?
    \item \textbf{RQ2}: What potential methods can mitigate the impact of degree distribution, thereby enhancing link prediction performance?
\end{itemize}

\subsubsection{Experimental results of stage-one KGE}
This study designs an experiment to analyze the differences in embedding quality between high- and low-degree entities in the FB15K-237-OWE knowledge graph dataset through longitudinal observation. Fig. 9 illustrates the embedding distributions of high- and low-degree entities. As shown, in the embedding space of high-degree entities, entities that are similar to the target entity (green points) are tightly clustered around the target entity (red point), while dissimilar entities (blue points) are situated further away. In contrast, such a clear distribution pattern is less evident in the embeddings of low-degree entities, resulting in a weaker discriminative ability of their embeddings. Visually, the embedding quality of high-degree entities is superior to that of low-degree entities, with the former's embedding distribution more closely aligning with the characteristics of high-quality embeddings.

\begin{figure}[htpb]
\centering
\resizebox{\textwidth}{!}{
\includegraphics[width=\textwidth, trim=0cm 0cm 0cm 0cm,clip]{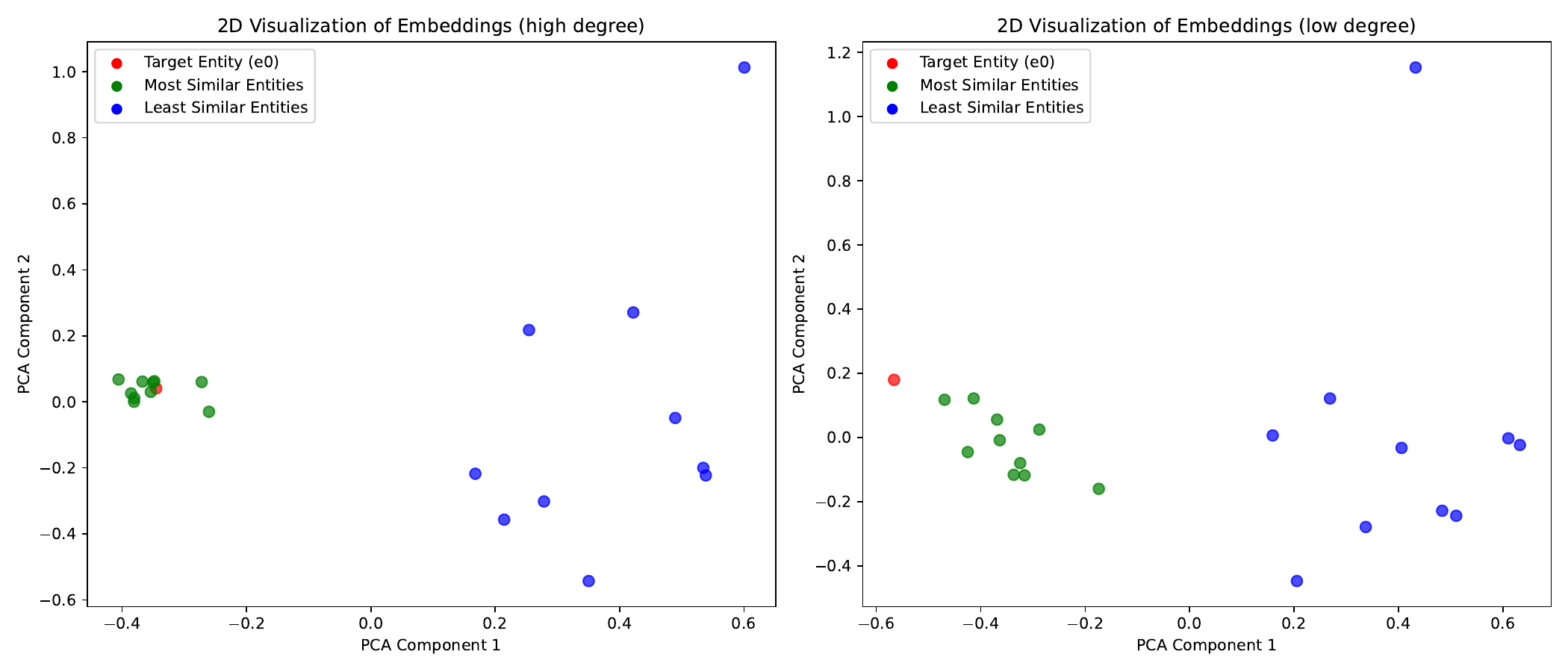}}
\caption{Visualization (partial) of the embedding space distribution for high- and low-degree entities. The left plot represents high-degree entities, and the right plot represents low-degree entities. In this figure, the red dots represent randomly selected high-degree entities, the green dots represent entities that are similar in terms of links, and the blue dots represent entities that are dissimilar.}\label{fig9}
\end{figure}

\begin{table}[]
\centering
\caption{Average embedding quality of high and low degree entities in FB15K237-OWE dataset }
\label{tab:4}
\resizebox{\textwidth}{!}{%

\begin{tabular}{llc}
\hline
\multicolumn{2}{l}{\multirow{2}{*}{}}                                          & \multirow{2}{*}{Average embedding quality index ${\bar Q}$} \\
\multicolumn{2}{l}{}                                                           &                                                  \\ \hline
\multicolumn{1}{c}{\multirow{2}{*}{FB15K237-OWE dataset}} & High degree entity & 0.535                                            \\ \cline{2-3} 
\multicolumn{1}{c}{}                                      & Low degree entity  & 0.166                                            \\ \hline
\end{tabular}
}

\end{table}

Table 4 summarizes the quantitative analysis of embedding quality for high- and low-degree entities. The experimental results reveal significant differences in embedding quality between entities of varying degrees, with high-degree entities achieving a higher average embedding quality index $\bar Q$ than low-degree entities. High-degree entities are associated with a greater number of triples in the knowledge graph and are sampled more frequently during training. This leads to more effective optimization of high-degree entities, enabling their embedding vectors to converge more easily to a stable global optimum. In contrast, low-degree entities are sampled less frequently, resulting in fewer updates to their embedding vectors, insufficient optimization, and lower embedding quality. The experimental findings reflect this disparity: link prediction performance is positively correlated with entity degree, with high-degree entities generally outperforming low-degree entities in prediction accuracy. In summary, entity degree directly influences its sampling frequency and embedding quality during KGE model training, indirectly affecting its performance in open-world link prediction tasks.

Furthermore, this study provides a cross-sectional analysis of KGE embedding quality under subgraphs with different degree distribution index. Table 5 presents the quantitative analysis results of embedding quality for subgraphs with high and low degree distribution index. The experimental results indicate that embedding space training quality differs under varying degree distributions. Subgraphs with high degree distribution index have a slightly lower overall average embedding quality index compared to those with low degree distribution index. However, focusing on high- and low-degree entities that play a critical role in overall link prediction performance, the average embedding quality index for these entities is higher in subgraphs with high degree distribution index than in those with low degree distribution index. This finding suggests that high- and low-degree entities are key to determining final link prediction performance. In subgraphs with high degree distribution index, knowledge graph links and triple training samples are more concentrated on certain high-degree entities. Since high-degree entities typically achieve better average embedding quality and MRR performance metrics, subgraphs with high degree distribution index deliver superior performance in open-world link prediction tasks.

\begin{table}[]
\centering
\caption{Average embedding quality of subgraphs with high and low degree distribution index }
\label{tab:5}
\resizebox{\textwidth}{!}{%
\begin{tabular}{ccc}
\hline
\multicolumn{2}{l}{\multirow{2}{*}{}}                                              & \multirow{2}{*}{Average embedding quality index} \\
\multicolumn{2}{l}{}                                                               &                                                  \\ \hline
\multirow{3}{*}{\parbox{5cm}{Subgraph with low \\ degree distribution index}}  & Low degree entity  & 0.153                                            \\
                                                              & High degree entity & 0.471                                            \\
                                                              & Total              & 0.392                                            \\ \hline
\multirow{3}{*}{\parbox{5cm}{Subgraph with high \\ degree distribution index}} & Low degree entity  & 0.172                                            \\
                                                              & High degree entity & 0.483                                            \\
                                                              & Total              & 0.375                                            \\ \hline
\end{tabular}
}
\end{table}

\subsubsection{Experimental results of Stage-Two mapping function training}

\begin{figure}[htpb]
\centering
\resizebox{\textwidth}{!}{
\includegraphics[width=\textwidth, trim=0cm 0cm 0cm 0cm,clip]{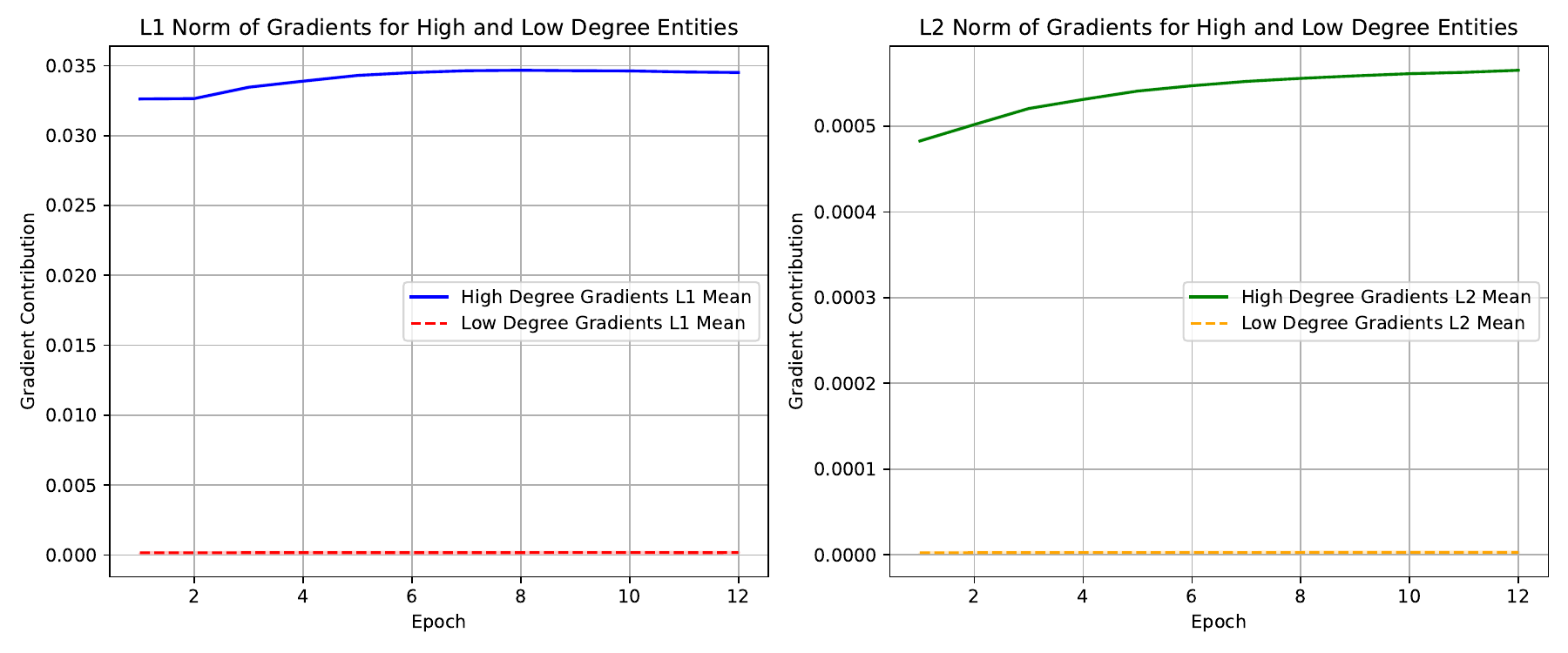}}
\caption{The curves showing the variation in gradient contributions of high- and low-degree entities across epochs. The left and right plots represent the results under L1 and L2 norms \citep{ref_44}, respectively.}\label{fig10}
\end{figure}

By categorizing entities into high- and low-degree groups and tracking the changes in their gradient contributions and alignment errors, we examine how entity degree influences the training of the mapping function. Fig.\ref{fig10} illustrates the variation in gradient contributions of high- and low-degree entities across epochs. The experimental results reveal that high-degree entities contribute significantly more to the gradient updates of neural network weights during training than low-degree entities (whether measured by L1 or L2 norms). This suggests that high-degree entities have a more pronounced effect on the model's weight updates during training. High-degree entities are sampled more frequently in the training process, leading to a larger cumulative impact of their gradients on weight updates. In contrast, low-degree entities are involved in fewer samples, resulting in limited gradient contributions and insufficient optimization.

\begin{figure}[htpb]
\centering
\resizebox{\textwidth}{!}{
\includegraphics[width=\textwidth, trim=0cm 0cm 0cm 0cm,clip]{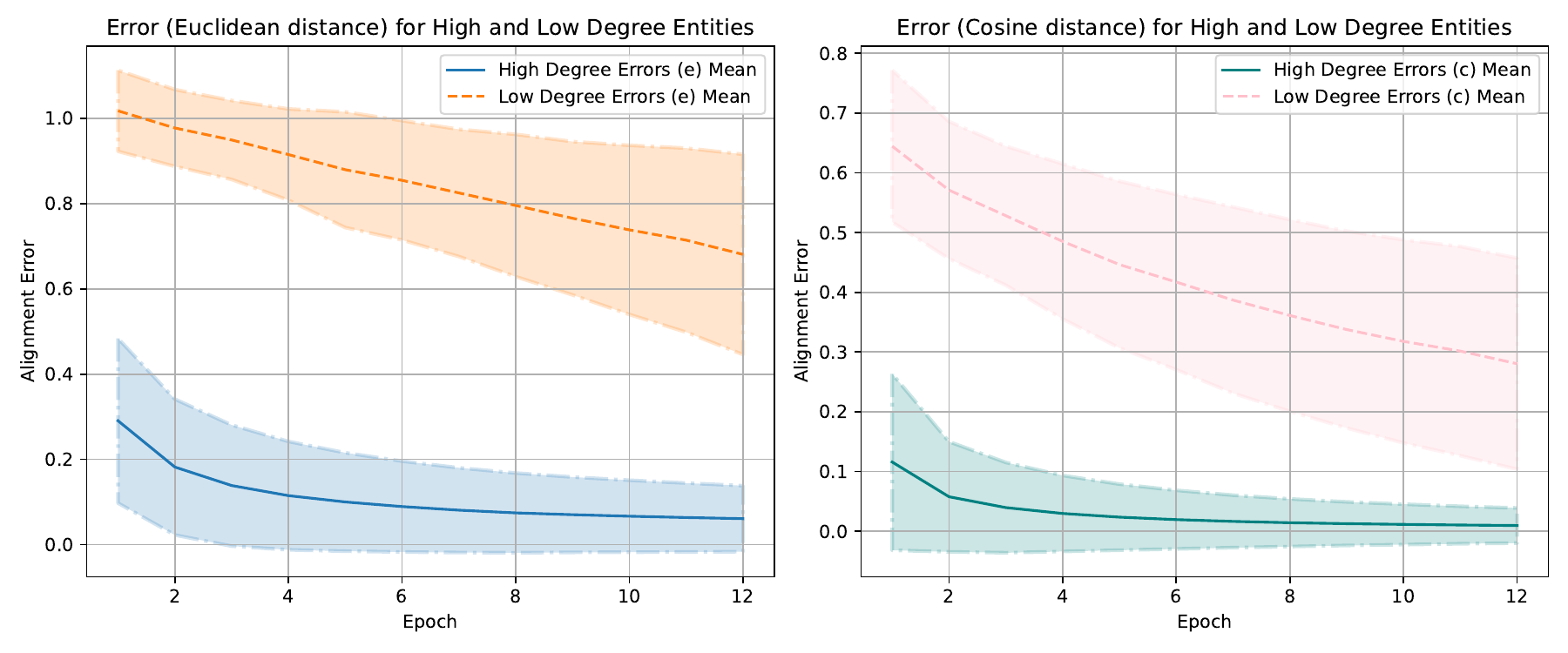}}
\caption{The changes in alignment error for high- and low-degree entities across epochs. The left and right plots represent the alignment error results calculated using Cosine distance and Euclidean distance, respectively.}\label{fig11}
\end{figure}

Fig.\ref{fig11} illustrates the changes in the alignment error and its standard deviation of high- and low-degree entities across epochs. As shown, the alignment error for high-degree entities rapidly converges to a lower value during training, while the standard deviation of the alignment error decreases progressively. In contrast, low-degree entities exhibit significantly higher alignment errors, with slower convergence and an increasing standard deviation as training continues. For high-degree entities, their graph embeddings are well-optimized during the first stage (KGE pretraining), making the alignment between text and graph embeddings relatively easier. Moreover, the higher sampling frequency of high-degree entities enables the mapping function to learn their alignment patterns more effectively, which facilitates faster convergence of the alignment error. The embedding quality of low-degree entities in the first stage also indirectly impacts the second-stage mapping function training. Instabilities in their semantic representation hinder the mapping function's ability to effectively capture their embedding features. Additionally, the limited number of training samples for low-degree entities restricts the learning of their alignment patterns, resulting in higher alignment errors. The increase in standard deviation further indicates that the model’s learning performance is inconsistent across different low-degree entities, leading to underfitting for some entities. Fig.12 shows the overall prediction performance (MRR) across epochs. It is evident that the model's performance peaks at the 11th epoch. At this point, the alignment error for low-degree entities has yet to converge, while that for high-degree entities has nearly converged, suggesting that further training may result in overfitting of high-degree entities. These experimental results highlight that the impact of entity degree on the mapping function training is both holistic and systemic. The underfitting of low-degree entities cannot be alleviated simply by increasing the number of training epochs, as this could lead to overfitting of high-degree entities, ultimately causing a decline in overall prediction performance.

\begin{figure}[htpb]
\centering
\resizebox{\textwidth}{!}{
\includegraphics[width=\textwidth, trim=0cm 0cm 0cm 0cm,clip]{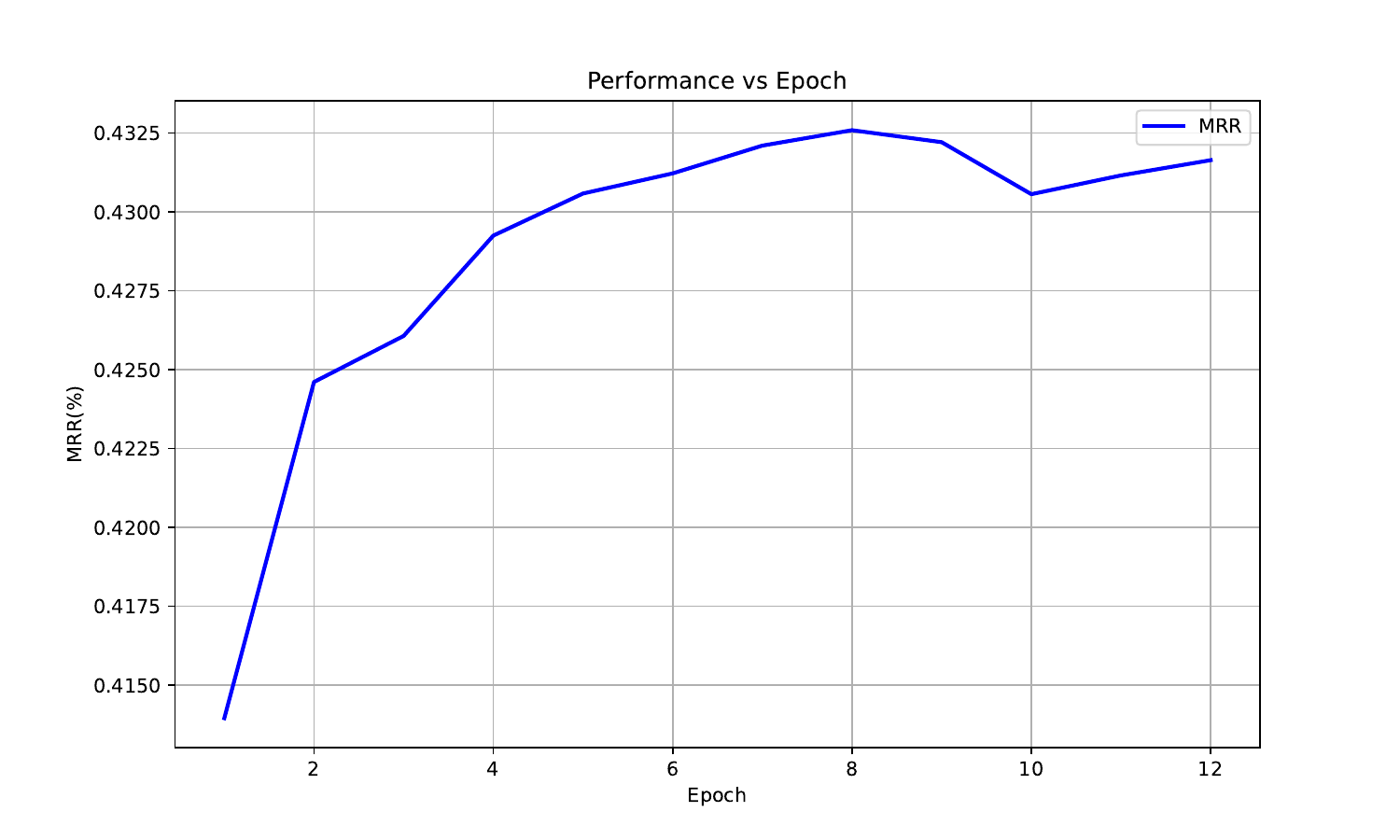}}
\caption{ The curve plot showing the variation in model prediction performance across training epochs.}\label{fig12}
\end{figure}

\subsubsection{Discussion on potential optimization approaches}
This study unveils the potential mechanisms through which entity degree influences the model training process. Based on these findings, future optimization strategies for open-world link prediction models can be pursued in several directions. First, data augmentation methods could be incorporated during stage one KGE training. In the embedding learning of low-degree entities, existing data augmentation techniques, such as Mix-up \citep{ref_45} and KG-Mixup \citep{ref_35}, could be employed. By weighted mixing the embeddings of similar entities with the embeddings of the low-degree entities themselves, richer training samples could be generated. Second, the sampling strategy in stage two mapping function alignment training can be optimized. To address the issue of a limited number of triple samples for low-degree entities, a sampling method based on entity degree distribution can be designed, increasing the proportion of low-degree entity samples. This would enable the mapping function to focus more on entities that are harder to align. Additionally, to tackle the larger alignment errors and slower convergence of low-degree entities, optimizing the loss function may be considered. These approaches are expected to mitigate the impact of degree distribution in the structural characteristics of knowledge graph datasets, helping to break through the current performance bottlenecks in open-world link prediction tasks.

%% Refer following link for more details.
%% https://en.wikibooks.org/wiki/LaTeX/Mathematics
%% https://en.wikibooks.org/wiki/LaTeX/Advanced_Mathematics

%% Use a table environment to create tables.
%% Refer following link for more details.
%% https://en.wikibooks.org/wiki/LaTeX/Tables

%% Use figure environment to create figures
%% Refer following link for more details.
%% https://en.wikibooks.org/wiki/LaTeX/Floats,_Figures_and_Captions

\section{Conclusion}\label{sec6}
This study examines the impact mechanism of degree distribution, a critical structural characteristic, on the performance of open-world link prediction (OW-LP). The experimental results reveal that degree distribution has a substantial effect on model performance, with both the degree distribution index and entity degree demonstrating a strong positive correlation with prediction accuracy. Furthermore, a quantitative analysis method is proposed to thoroughly investigate the role of degree distribution in training OW-LP models. The findings show that degree distribution influences overall model performance by affecting both the quality of the embedding space and the gradient contributions during training. Specifically, high-degree entities are learned significantly better than low-degree entities, whose insufficient learning becomes a limiting factor in improving model performance. Based on these insights, this study discusses several potential optimization strategies, including enhancing sampling methods, incorporating data augmentation techniques, and designing adaptive loss functions, to improve the model's overall link prediction performance. In conclusion, this work not only confirms the critical role of degree distribution in open-world link prediction performance but also sheds light on its underlying mechanisms, providing valuable insights for future research in this field.
\section*{Acknowledgements}
This work was supported by the National Natural Science Foundation of China under Grant U1801262 and the Science and Technology Project of Guangdong Province under Grant 2019B010154003.

\bibliographystyle{model5-names} 
\bibliography{ref}



\end{document}